\documentclass[%
 reprint,
 amsmath,amssymb,
 aps,
]{revtex4-2}

\usepackage{graphicx}
\usepackage{dcolumn}
\usepackage{bm}
\usepackage[mathlines]{lineno}
\usepackage{tabularx}
\usepackage{booktabs}
\usepackage{array}
\usepackage{makecell}

\begin{document}

\preprint{APS/123-QED}

\title{Percolation of a cohesive fine particle in a static bed}

\author{
Jizhi Zhang$^{1}$,
Qiong Zhang$^{2}$,
Julio M.~Ottino$^{2,3,4}$,
Paul B.~Umbanhowar$^{2}$,
and Richard M.~Lueptow$^{2,3,4}$
}
\email{r-lueptow@northwestern.edu}

\affiliation{$^{1}$Theoretical and Applied Mechanics Program, Northwestern University, Evanston, Illinois 60208, USA}
\affiliation{$^{2}$Department of Mechanical Engineering, Northwestern University, Evanston, Illinois 60208, USA}
\affiliation{$^{3}$Department of Chemical and Biological Engineering, Northwestern University, Evanston, Illinois 60208, USA}
\affiliation{$^{4}$Northwestern Institute on Complex Systems (NICO), Northwestern University, Evanston, Illinois 60208, USA}

\date{\today}

\begin{abstract}
Percolation of fine particles (fines) in a static bed of larger particles is central to many industrial and natural processes. Non-cohesive fines either pass through the bed or become trapped depending on multiple factors including particle sizes, friction and restitution coefficients, and size-polydispersity. Here we consider the additional factor of cohesion. We use the discrete element method to simulate gravity-driven percolation of cohesive fine particles through a static bed of randomly packed large particles; fines interact with bed particles but not with each other. A large-to-fine particle diameter ratio of 7 geometrically permits non-cohesive fines to pass the narrowest pore throats formed by the large particles so they can freely percolate. However, sufficiently large cohesion and friction lead to non-geometric trapping. Fines are trapped when they fail to rebound after a collision, due to large cohesion, low restitution, and low collision velocity, and any subsequent rolling or sliding is insufficient to cause detachment. This establishes a sequence of local interactions -- collision, adhesion, and post-contact motion -- that governs the ultimate fate of a fine particle. A collisional model that incorporates a trapping probability per collision and a collision frequency predicts the trapping distance in the regime dominated by collision-induced trapping. For non-rebounding collisions, frictional effects are enhanced by cohesion and, when large enough, prevent the fine particle from subsequently detaching. A static equilibrium condition based on force balance predicts whether a fine particle remains stationary after contact. These results show that percolation of cohesive fine particles is not determined by geometric accessibility alone, but also by particle-scale interaction dynamics that can override geometric expectations.

\end{abstract}

\maketitle


\section{Introduction}

Understanding the behavior of granular materials is essential for a wide range of industrial applications, including pharmaceutical manufacturing and chemical processing, as well as a variety of geophysical phenomena~\cite{Jaeger1996,Duran2012}. Fine particles, or fines, are common in industrial solids handling processes due to the inherent polydispersity of most raw materials, particle attrition and comminution, or the intentional addition of particles such as free-flow agents. However, the presence of fines can lead to undesirable outcomes including degraded product quality, equipment fouling, health risks from inhalation, and safety hazards such as dust explosions~\cite{Bemrose1987,Schulze2008}. Beyond industrial applications, fine particles play a central role in geophysical flows. In environmental and subsurface settings, the migration of fines through coarser granular matrices governs a range of critical processes, including riverine sediment transport, contaminant migration in groundwater, and clogging in filtration systems~\cite{sharma1987,Fang2024}. When fine particles are much smaller than surrounding large particles, they can easily pass through interstices between the large particles to percolate downward due to gravity, a process called free sifting. In a particle bed composed of monodisperse spheres, free-sifting of fines occurs when the large particle to fine particle diameter ratio, $R$, is larger than the geometric trapping threshold, $R_t = d_l/d_f = \sqrt{3}/(2 - \sqrt{3}) \approx 6.464$, where $d_l$ is the large particle diameter and $d_f$ is the fine particle diameter, corresponding to the smallest pore throat formed by three contacting bed spheres~\cite{Ippolito2000,Lomine2009}, although polydispersity or particle deformation can increase $R_t$~\cite{Vyas2024}.

Previous studies in static beds of large particles find that an untrapped fine particle percolates downward with a constant average velocity while exhibiting additional diffusive motion both perpendicular and parallel to the direction of gravity~\cite{Bridgwater1969,Bridgwater1971,Masliyah1974,Gao2023,Vyas2024,Vyas2025b}. Its percolation velocity varies inversely with its root mean square (rms) velocity, which depends strongly on $R$ and coefficients of restitution and friction~\cite{Vyas2025b}. Beyond the single-particle limit, fine particle concentration is critical. At low concentration, the fine particles percolate through the static bed independently of one another, while high concentration of fine particles can lead to clogging within pore throats, thereby reducing percolation~\cite{Lomine2009,Remond2010}.

While geometric constraints determine whether a fine particle can physically pass through a pore throat, dynamic interactions like cohesion can also play a crucial role in trapping and percolation behavior. Cohesive forces arise from a variety of physical mechanisms~\cite{Sharma2025}, including semi-permanent solid bonding between particles~\cite{Iveson1996}, van der Waals and electrostatic interactions~\cite{Visser1989}, and liquid capillary bridges~\cite{Bocquet1998,Halsey1998}. In flowing granular materials, the presence of cohesion alters contact behavior between particles, including friction~\cite{Marshall2009,Krijt2014} and restitution~\cite{Abbasfard2016,Murphy2017,Wang2025}, which strongly influences macroscopic flow properties. As a result, the flow of cohesive materials deviates markedly from that of non-cohesive systems. Cohesive interactions promote enduring contacts and agglomeration that modify particle mobility and local rearrangement dynamics~\cite{Castellanos2001,Tomas2004}. Continuum descriptions of flows of monodisperse and polydisperse cohesive granular materials use population‑balance approaches and contact‑based models~\cite{Kellogg2017,Kellogg2023}, and there has been extensive work on the rheology of cohesive granular flows~\cite{Mandal2020,Mandal2021}. However, these frameworks offer limited physical insights into systems for which the size ratio exceeds the geometric trapping threshold, where percolation mechanisms change qualitatively, leaving a gap in understanding how cohesion influences percolation in such systems.

In this study, we use discrete element method (DEM) simulations to investigate the influence of cohesion on the percolation of cohesive fine particles through randomly packed static beds of larger particles in the single particle limit (i.e., fines interact with bed particles but not with each other) for $R=7>R_t$ (Fig.~\ref{fig:geometry}(a)). By systematically varying the cohesion, restitution, and friction, we characterize their effects on fine particle percolation and trapping. Our results reveal that even in the free-sifting regime, fine particles can become immobilized due to the combined effects of cohesion, restitution, and friction, and that percolation dynamics are governed not only by geometry but also by local interactions and bed structure. These findings provide a foundation for understanding the more difficult problem of granular segregation in flowing granular mixtures of cohesive particles at large particle size ratios.

The remainder of the paper is organized as follows. Section~\ref{sec:Methods} describes the numerical simulations and provides an overview of the cohesive contact models. Section~\ref{sec:Results} presents results for the percolation behavior in both the passing and trapping regimes. We introduce the inverse trapping distance as a key metric to characterize trapping and systematically consider the effects of cohesion, restitution coefficient, friction, and Young's modulus. A model is developed to capture the coupled effects of cohesion, restitution, and impact velocity, which together govern whether a fine particle rebounds after colliding with a bed particle. Additionally, we identify a static equilibrium condition that distinguishes different post-collision behaviors for non-rebounding fine particles. We also briefly discuss the case where a single fine particle simultaneously contacts two large particles. Section~\ref{sec:Conclusions} summarizes our findings.

\section{Methods} \label{sec:Methods}

\subsection{Numerical simulations}\label{sec:simulation}
To study the percolation of cohesive fine particles in a static bed, we implement a standard soft-sphere DEM approach using the open-source simulation software LAMMPS~\cite{Plimpton1995,Thompson2022}. Numerical simulation is performed using an improved Velocity-Verlet scheme developed for systems with large size ratios~\cite{Vyas2025a}. Normal contact forces follow the Hertzian elastic model~\cite{Thornton2011} with a Young’s modulus of $E=10^{7}\,\mathrm{Pa}$, unless otherwise noted, and a Poisson ratio of $\nu = 0.17$. Inelasticity is incorporated using a velocity-independent constant-restitution damping model~\cite{Brilliantov1996,Vyas2025a} with normal restitution coefficient $e_n=0.8$ unless otherwise noted. However, the restitution coefficient is only independent of the impact velocity for non-cohesive particles. Tangential forces are described by the Mindlin model~\cite{Mindlin1949} with a sliding friction coefficient $\mu$ varied between $0$ and $0.5$, coupled with the Marshall twisting resistance~\cite{Marshall2009}. Rolling resistance is included using the spring-dashpot-slider model~\cite{Luding2008,Wang2015} with rolling coefficient $\mu_r=0.5$ to suppress rolling behavior and emphasize sliding behavior. The particle density is $\rho=2500\,\mathrm{kg/m^3}$ for all particles. The large particle diameter is fixed at $d_l = 4\,\mathrm{mm}$, and the large-to-fine particle diameter ratio is $R=7$, corresponding to a fine particle diameter of $d_f \approx 0.57\,\mathrm{mm}$. These parameters ensure maximum particle overlaps smaller than 1\% of the fine particle diameter~\cite{Vyas2025a}, which is sufficient for the dilute, collision‑dominated regime considered here, while maintaining moderate computational cost. To fully resolve collision dynamics, the DEM timestep is one-tenth of the characteristic collision time, resulting in a timestep of approximately $10^{-5}\,\mathrm{s}$. Convergence tests confirm that this choice of timestep is sufficient to resolve the collisions (i.e., smaller timesteps do not alter the simulation results).

\begin{figure}[htbp]
  \centering
  \includegraphics[width=1\linewidth]{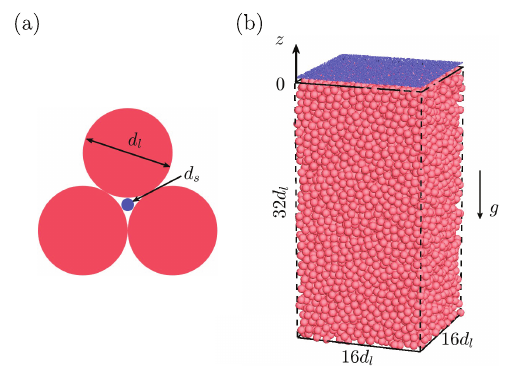}
  \caption{(a) Example of a pore throat containing a fine particle for a large-to-fine diameter ratio of $R=7$ at virtual overlap $\delta=0$. (b) Perspective view of the static bed of large particles (red) with packing fraction $\phi = 0.6$ and periodic vertical boundaries, showing the initial positions of the non-interacting percolating fines (blue).}
  \label{fig:geometry}
\end{figure}

The computational domain is a random packing of spheres in a $16d_l\times 16d_l \times 32 d_l$ rectangular prism as shown in Fig.~\ref{fig:geometry}(b). The large particles are monodisperse and identical to avoid changes to $R_t$ due to polydispersity~\cite{Vyas2024}. To avoid wall effects, vertical boundaries of the box are periodic, while the non-periodic top and bottom boundaries allow fine particles to enter the bed at the top and exit the bed at the bottom. A particle growth algorithm is used to form the bed and ensure that bed particles do not unphysically overlap~\cite{Gao2023}, after which the large particles are immobilized to fix the bed structure. The bed packing fraction is set to $\phi=0.6$, which requires 9387 large particles.

To initialize each simulation, fine particles are randomly positioned in a horizontal plane located a distance $d_f + d_l$ above the center of the highest bed particle, shown in Fig.~\ref{fig:geometry}(b). After release, the fine particles fall from rest under gravity, $g = 9.8\,\mathrm{m/s^2}$. The release height does not influence the subsequent kinematics once a particle descends $2d_l$ into the bed. Because the focus here is on the dynamics of single fine particles, fine particles interact only with large particles and not with one another. The dynamics of $10^4$ independent intruder fine particles are simultaneously computed in each simulation, which is computationally efficient and provides sufficient data for accurate statistics.

\subsection{Cohesive contact model}\label{sec:model}
We characterize the strength of the cohesive interaction between particles by comparing it to the particle weight. For any cohesion mechanism, a dimensionless Bond number can be defined as~\cite{Nase2001,Sharma2025}
\begin{equation} \label{eq:Bond_numner}
    Bo = \frac{F_c}{W},
\end{equation}
where $F_c$ denotes the pull-off force, i.e., the minimum force needed to separate contacting particles, and $W$ denotes the particle weight. Note that this definition is the inverse of the Bond number commonly used in the context of capillary flow~\cite{De2003}. Some studies may use other parameters to characterize cohesion, but the underlying physical interpretation remains consistent~\cite{Mandal2020,Ralaiarisoa2022}. Since we are interested in the kinematics of the fine particles, we use the weight of the fine particle, $W=m_fg$, rather than a form for $Bo$ that includes the mass of both particles, as is sometimes done for mixtures of bidisperse cohesive particles~\cite{Jarray2019}.

Given the diversity of cohesion mechanisms, various models can be employed to represent cohesive contacts between particles. Table~\ref{tab:cohesive_models_compare} summarizes normal contact forces and pull-off forces for the widely used DMT~\cite{Derjaguin1975} and JKR~\cite{Johnson1971} cohesive contact models, and compares them to the non-cohesive Hertzian contact model~\cite{Thornton2011}. In these expressions, $\sigma$ is the surface energy density, $E_{\mathrm{eff}}=E/2(1-\nu^2)$ represents the effective stiffness, $r_{\text{eff}}= d_l  d_f/2(d_l+d_f)$ is the effective particle radius, $\delta$ is the virtual overlap between contacting particles, and $a$ is the contact patch radius, obtained numerically from $\delta = a^2/r_{\text{eff}} - 2\sqrt{\pi \sigma a/E_{\mathrm{eff}}}$.

\begin{table}[h]
\centering
\renewcommand{\arraystretch}{1.5}
\setlength{\belowcaptionskip}{10pt}
\caption{Contact models used in this work.}
\begin{tabularx}{\columnwidth}{c @{\hspace{16pt}} c @{\hspace{0pt}} c}
\toprule
Model & Contact force & Pull-off force, $F_c$ \\
\midrule\midrule
Hertz &
$\displaystyle F_{\mathrm{Hertz}} = \frac{4}{3} E_{\mathrm{eff}} r_{\mathrm{eff}}^{1/2} \delta^{3/2}$ &
$0$ \\
DMT &
$\displaystyle F_{\mathrm{DMT}} = \frac{4}{3} E_{\mathrm{eff}} r_{\mathrm{eff}}^{1/2} \delta^{3/2} - 4 \pi \sigma r_{\mathrm{eff}}$ &
$\displaystyle 4 \pi \sigma r_{\mathrm{eff}}$ \\
JKR &
$\displaystyle F_{\mathrm{JKR}} = \frac{4 E_{\mathrm{eff}} a^{3}}{3 r_{\mathrm{eff}}} - 2 \pi a^{2} 
\sqrt{\frac{4 \sigma E_{\mathrm{eff}}}{\pi a}}$ &
$\displaystyle 3 \pi \sigma r_{\mathrm{eff}}$ \\
\bottomrule
\end{tabularx}
\label{tab:cohesive_models_compare}
\end{table}

\begin{figure}[htbp]
  \centering
  \includegraphics[width=1\linewidth]{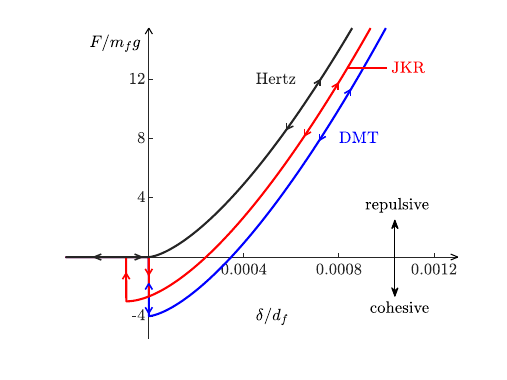}
  \caption{Comparison of $\delta$-dependent normal forces for cohesive (JKR, DMT) and non-cohesive (Hertz) contact models for surface energy density $\sigma = 0.003\, \mathrm{J/m^2}$; see Table~\ref{tab:cohesive_models_compare}.}
  \label{fig:static_cohesion}
\end{figure}

Cohesive particles experience attractive forces upon initial contact at $\delta=0$. As $\delta$ increases, the repulsive force related to particle stiffness and deformation grows stronger and opposes the attractive cohesive force. The detailed behavior of these interactions differs between the cohesive contact models, as illustrated in Fig.~\ref{fig:static_cohesion}: $F>0$ indicates a repulsive net force when particle stiffness dominates, and $F<0$ indicates a net attractive force when cohesion dominates. The most negative value for $F$ corresponds to the pull-off force, $F_c$. For initial contact at $\delta = 0$, the DMT and JKR models exhibit a normal force discontinuity. Additionally, the JKR model displays hysteresis during particle separation, i.e. $F<0$ for $\delta/d_f < 0$. Viscous damping during contact is included separately as discussed in Sec.~\ref{sec:simulation}.

In this study we primarily use the DMT model for which the cohesive force during contact is constant and the repulsive contact force is Hertzian. The DMT model is simpler and has no energy loss resulting from the hysteretic contact force related to the contact patch radius, $a$, in the JKR model. This facilitates a clearer interpretation of the underlying physics. Although the present parameters do not strictly satisfy the DMT assumption of small contact deformation~\cite{Tabor1977}, the model nevertheless captures the essential interaction mechanisms relevant to the system. We  also perform a limited number of simulations using the JKR model, which yield qualitatively similar percolation behavior.

\section{Results} \label{sec:Results}

\subsection{Passing and trapping}

\begin{figure}[htbp]
  \centering
  {%
    \includegraphics[width=1\linewidth]{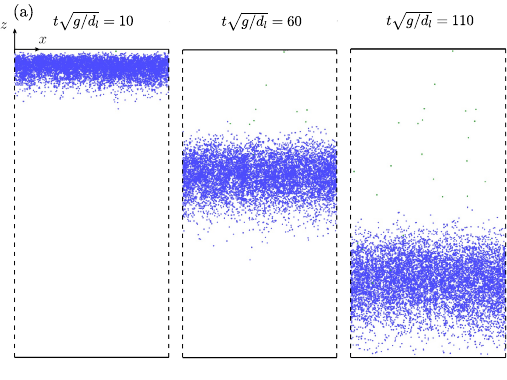}
  }\\[1em]
  {%
    \includegraphics[width=1\linewidth]{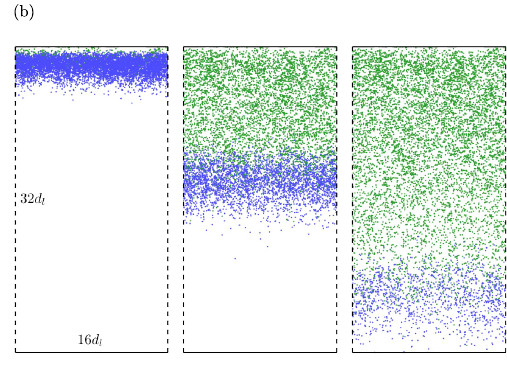}
  }
  \caption{Positions of $10^4$ non-interacting fine particles in a static bed with $\mu = 0.01$ and $R=7$ at different times (columns) with (a) $Bo=0$, (b) $Bo=16$. Large particles are omitted for clarity. Percolating fines are colored blue while trapped fines are colored green.}
  \label{fig:passing_and_trapping}
\end{figure}

We begin by illustrating passing and trapping behavior of fine-particle percolation in a static bed of large particles at $R=7$ for varying levels of cohesion in Fig.~\ref{fig:passing_and_trapping}. The non-interacting fine particles enter at the top and percolate through the static bed until they are trapped or exit the domain upon reaching the bottom. In the non-cohesive case, fine particles pass through the bed because their diameter is less than the smallest possible pore-throat diameter. Under these conditions, fine particles percolate downward at similar velocities, with only a few becoming immobilized (trapped) due to frictional forces at double contacts~\cite{Vyas2025a}. In contrast, strong cohesion promotes trapping behavior even at $R=7$, which is above the geometric trapping threshold of $R_t \approx 6.464$. As time progresses, more fine particles become trapped in the bed so that fewer particles percolate downward to the lower regions of the bed.

\begin{figure}[htbp]
  \centering
  \includegraphics[width=1\linewidth]{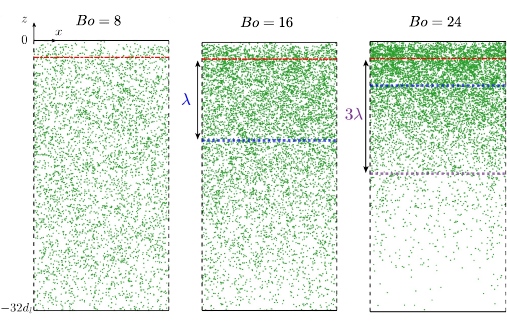}
  \caption{Positions of trapped fine particles for different cohesion levels ($Bo$) with $\mu=0.01$ and $R=7$ after $t\sqrt{g/d_l}=500$. Red dashed-dotted lines denote $z_0=-2d_l$ below which statistics are obtained for trapped particles. Blue dotted lines denote the characteristic distance $\lambda$ below $z_0=-2d_l$, over which 63\% of the remaining untrapped particles become trapped, and the purple dotted line indicates $3\lambda$, corresponding to 95\% trapping (lines outside the simulation region are omitted).}
  \label{fig:percolattion}
\end{figure}

The final positions after all particles have either passed out the bottom of the domain or become trapped, are shown in Fig.~\ref{fig:percolattion} for different cohesion levels at dimensionless
time $t\sqrt{g/d_l}=500$, after all motion has ended, corresponding to the end of the simulation. As $Bo$ increases, particles become trapped more quickly after falling shorter distances, resulting in shallow penetration. These results clearly demonstrate the strong influence of cohesion on passing and trapping behavior.

\subsection{Trapping distribution and trapping distance}
\begin{figure}[htbp]
  \centering
  \includegraphics[width=1\linewidth]{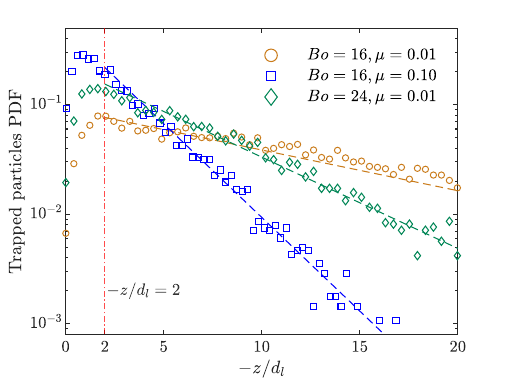}
  \caption{Examples of trapped particle distributions (PDF) and exponential fits (dashed lines) for $-z/d_l>2$ (vertical dashed-dotted line).}
  \label{fig:exponential_distribution}
\end{figure}

To characterize the trapping regime, we consider the probability density function (PDF) for the vertical positions of trapped fine particles, shown in Fig.~\ref{fig:exponential_distribution}. Because the first two layers of large particles act as a transition region from freely falling fine particles to steady percolation dynamics, we only consider particles trapped below $z_0 = -2d_l$. Figure~\ref{fig:exponential_distribution} also shows exponential fits (dashed lines) to the PDF of the trapped-particle distributions below this height. Since the trapping statistics follow an exponential form, we characterize the probability for vertical trapping position of fine particles as
\begin{equation}\label{eq:exponential}
    \mathrm{PDF}(z) = \frac{1}{\lambda} e^{{\Delta z}/\lambda},\,\, -\infty < z \leq z_0,
\end{equation}
where $\Delta z = z-z_0$ and $\lambda$ is the characteristic percolation distance beyond $z_0=-2d_l$ based on the exponential fit. Physically, $\lambda$ is the characteristic distance below $z_0=-2d_l$ over which 63\% of the remaining untrapped particles become trapped; 95\% become trapped in a distance $3 \lambda$. For instance, for $Bo=24$ and $\mu=0.01$ in Fig.~\ref{fig:exponential_distribution}, $\lambda = 5.3d_l$, meaning that 63\% of the remaining untrapped particles at $z_0=-2d_l$ become trapped by $z = -7.3 d_l$ and 95\% become trapped by $z = -17.9 d_l$, shown as horizontal dotted lines for the $Bo=24$ case in Fig.~\ref{fig:percolattion}.

\subsection{Influence of parameters on trapping}

Several factors combine with cohesion to play significant roles in trapping behavior, including size ratio, sliding friction, rolling friction, Young’s modulus, and restitution coefficient. Since previous work has addressed the effects of size ratio~\cite{Gao2023} and restitution coefficient~\cite{Vyas2025b}, we focus here on the influence of cohesion, sliding friction, and particle stiffness.

\subsubsection{Effects of cohesion and sliding friction}

\begin{figure}[htbp]
  \centering
  \includegraphics[width=1\linewidth]{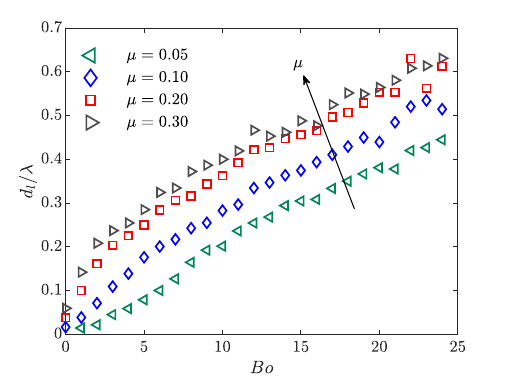}
  \caption{Inverse trapping distance, $d_l/\lambda$, vs $Bo$ for sliding friction coefficients $\mu \in\{0.05,\,0.10, \,0.20,\,0.30\}$.}
  \label{fig:trapping_length_Bo}
\end{figure}

Consider first the influence of cohesion, quantified by the Bond number $Bo$, and sliding friction $\mu$ on trapping of fine particles that would otherwise sift freely through the packed bed at $R=7$. Figure~\ref{fig:trapping_length_Bo} shows the dimensionless inverse trapping distance, $d_l/\lambda$, as a function of $Bo$ for sliding friction coefficients $\mu \in\{0.05,\,0.10,\,0.20,\,0.30\}$. Percolation with no trapping corresponds to $d_l/\lambda \rightarrow 0$, while shallow penetration corresponds to large $d_l/\lambda$. In the non-cohesive limit, trapping results from frictional contact alone, amplified by large rolling resistance~\cite{Vyas2025b}. In all cases, the inverse trapping distance increases monotonically with $Bo$. For $Bo \lesssim 6$, both cohesion and sliding friction play roles in trapping, with $d_l/\lambda \rightarrow 0$ (no trapping) as cohesion and friction approach 0. For $Bo \gtrsim 6 $ the dependence of $d_l/\lambda$ on $Bo$ is approximately linear and the slopes for different friction levels are similar, indicating that cohesion governs the trapping behavior. As $Bo$ increases further, most particles become trapped within $\Delta z \approx -d_l$, rendering statistics for $\lambda$ unreliable. Along similar lines, for all of the cases in Fig.~\ref{fig:trapping_length_Bo}, the distance for 95\% of the free particles to be trapped exceeds $5 d_l$. As friction increases, the curves increasingly overlap, suggesting that the trapping behavior saturates and becomes insensitive to further increases in sliding friction.

\begin{figure}[htbp]
  \centering
  \includegraphics[width=1\linewidth]{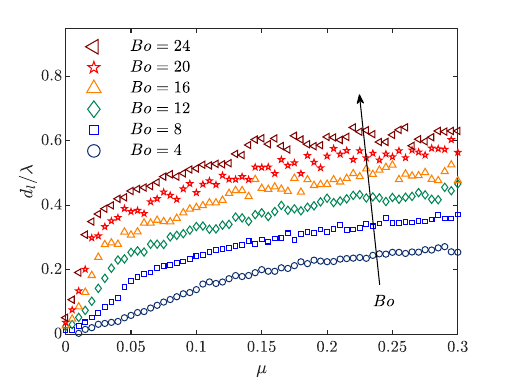}
  \caption{Inverse trapping distance, $d_l/\lambda$, vs sliding friction, $\mu$, for $4 \le Bo \le 24$.}
  \label{fig:trapping_length_friction}
\end{figure}

Replotting this data along with results for additional values of $\mu$ further reveals the influence of sliding friction on trapping behavior, as shown in Fig.~\ref{fig:trapping_length_friction}. For all values of $Bo$, the inverse trapping distance increases with $\mu$ and exhibits a kink at low $\mu$ that suggests two regimes. At small $\mu$, the slope is relatively steep, indicating strong dependence on $\mu$. At larger $\mu$ the slope becomes shallow and the curves flatten out as cohesion becomes the dominant trapping mechanism. For $\mu$ values greater than the range shown ($\mu>0.3$), the inverse trapping distance is constant, consistent with the saturation observed in Fig.~\ref{fig:trapping_length_Bo}. Note that the position of the kink shifts toward smaller friction coefficients as $Bo$ increases, an effect that is examined in Sec.~\ref{sec:friction}.

\begin{figure}[htbp]
  \centering
  \includegraphics[width=1\linewidth]{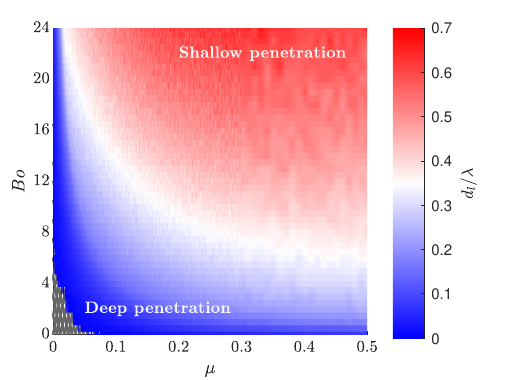}
  \caption{Inverse trapping distance, $d_l/\lambda$, dependence on cohesion, $Bo$, and sliding friction, $\mu$. Hashed pattern indicates conditions under which a negligible number of fines are trapped.}
  \label{fig:phase_chart}
\end{figure}

The combined influence of cohesion and sliding friction on $d_l/\lambda$ is shown in Fig.~\ref{fig:phase_chart}. In the upper-right region of the figure, where both friction and cohesion forces are strong, $d_l/\lambda$ is large, indicating shallow penetration. As friction and cohesion are decreased, $d_l/\lambda$ decreases, corresponding to deeper penetration. When both $\mu$ and $Bo$ are small (near the origin), most particles undergo free sifting (hatched), resulting in negligible trapping. For $\mu>0.5$ (not shown in Fig.~\ref{fig:phase_chart}), $d_l/\lambda$ is effectively constant for all $Bo \gtrapprox1$. For $\mu>0.1$ and $Bo>24$, nearly all particles are immediately trapped upon entering the bed. Finally, reducing the rolling resistance increases the trapping distance, but does not modify the qualitative trends.

As noted with respect to Figs.~\ref{fig:trapping_length_Bo} and~\ref{fig:trapping_length_friction}, strong dependence on $\mu$ is evident at low $Bo$ and strong dependence on $Bo$ is evident at large $\mu$. Regions with the same color in Fig.~\ref{fig:phase_chart} correspond to equivalent trapping distances, which indicates similar trapping behavior for various parameter combinations. Hence, trapping behavior is not governed by cohesion alone; sliding friction also plays a key role, particularly for $\mu \lessapprox0.25$.

\subsubsection{Effects of particle stiffness}

\begin{figure}[htbp]
  \centering
  \includegraphics[width=1\linewidth]{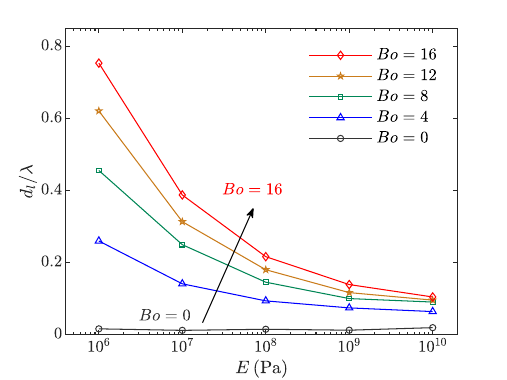}
  \caption{Inverse trapping distance, $d_l/\lambda$, vs Young’s modulus, $E$, for $Bo \in \{0,4,8,12,16\}$.}
  \label{fig:trapping_length_YoungsM}
\end{figure}

Particle stiffness, as controlled by the Young's Modulus, $E$, can play a significant role in cohesive trapping behavior. While the Hertzian contact force depends directly on $E$, the cohesive force in the DMT model used here is unaffected by particle stiffness. To consider the impact of stiffness, we simultaneously change $E$ for both the large bed particles and percolating fine particles. Since particle stiffness can affect the packing fraction, which in turn affects particle percolation~\cite{Vyas2024}, the bed packing fraction is kept constant at $\phi=0.6$ to isolate the effects of particle stiffness on percolation. As a baseline, in non-cohesive systems, stiffness has negligible influence on trapping, as would be expected for a free sifting size ratio of $R=7$, as shown in Fig.~\ref{fig:trapping_length_YoungsM} for $Bo=0$, although a small number of fine particles become trapped due to friction~\cite{Vyas2025a} leading to a non-zero value of $d_l/\lambda$. When particles are cohesive, softer particles (lower $E$) are more easily trapped, reducing the trapping distance $\lambda$, particularly for larger $Bo$.
The difference in inverse trapping distance between different cohesion levels diminishes as particles become stiffer.

\subsubsection{Comparing $\mathrm{DMT}$ and $\mathrm{JKR}$ models}

\begin{figure}[h!]
  \centering
  \includegraphics[width=1\linewidth]{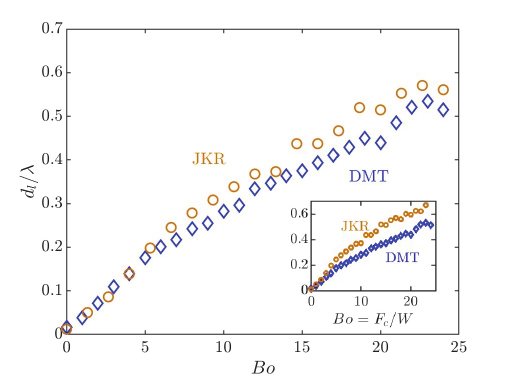}
  \caption{Inverse trapping distances, $d_l/\lambda$, vs $Bo_{\mathrm{DMT}} = 4 \pi \sigma r_{\mathrm{eff}}/W$ for $\mu = 0.1$ for DMT and JKR models. Inset compares the models using the definition for $Bo$ in Eq.~\eqref{eq:Bond_numner}.}
  \label{fig:trapping_length_Bo_jkr}
\end{figure}

To test the sensitivity of our results to the cohesion model, we compare the inverse trapping distance for the DMT and JKR models. Because the Bond number based on the pull-off force in Eq.~\eqref{eq:Bond_numner} is different for the DMT and JKR models for the same surface energy density (see Table~\ref{tab:cohesive_models_compare}), that is, $Bo_{\mathrm{JKR}}=0.75 \,Bo_{\mathrm{DMT}}$, we compare the two models using equivalent Bond numbers based on the surface energy density (that is, $Bo:=Bo_{\mathrm{DMT}}=4 \pi \sigma r_{\mathrm{eff}}/W$).  This results in approximate collapse for the two models, as shown in Fig.~\ref{fig:trapping_length_Bo_jkr} for $\mu=0.1$. In contrast, when using the Bond number defined based on $F_c$, which is shown in the inset of Fig.~\ref{fig:trapping_length_Bo_jkr}, the JKR model yields a larger value for $d_l/\lambda$ than the DMT model, corresponding to shallower penetration. 

This behavior can be understood from an energy-based perspective. The distinction between the two cohesive contact models arises primarily from the initial collision outcome, whether rebounding occurs or not, whereas the subsequent post-contact motion, once rebounding is suppressed, depends predominantly on the force level. When the surface energy density is the same for the two cohesive contact models, the inelastic energy loss associated with normal damping is nearly identical. As a result, the trapping behavior in the collision-dominated regime becomes energetically equivalent, leading to a near overlap of the DMT and JKR results using $Bo:=Bo_{\mathrm{DMT}}=4 \pi \sigma r_{\mathrm{eff}}/W$. This indicates that, when collisions govern the trapping process, the surface energy density rather than the pull-off force provides the appropriate control parameter, and the Bond number must be modified accordingly to reflect this equivalence. Even so, the JKR model yields a slightly larger inverse trapping distance, corresponding to shallower penetration. This difference occurs because the JKR model includes additional energy loss due to hysteresis for separations when $\delta < 0$, shown in Fig.~\ref{fig:static_cohesion}. However, this energy loss is small compared with the energy dissipation from inelastic damping. As a result, both models exhibit qualitatively similar trends, including a decrease of the slope above $Bo \approx 6$ where $d_l/\lambda$ depends linearly on $Bo$.

\subsection{Two-Particle Interactions}

\begin{figure}[htbp]
  \centering
  \includegraphics[width=1\linewidth]{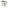}
  \caption{Examples of the four possible types of post-impact behavior. Two result in trapping ((b) sticking; (c) sliding \& trapping) and two in continued percolation ((a) rebounding; (d) sliding \& detaching). Parameters are: (a) $Bo=8, e_n=0.8, \mu=0.3$, (b) $Bo=8, e_n=0.2, \mu=0.3$, (c) $Bo=8, e_n=0.2, \mu=0.02$, (d) $Bo=3, e_n=0.2, \mu=0.02$.}
  \label{fig:two_particle_interactions}
\end{figure}

To establish a foundation for understanding the observed trapping behavior, we begin by considering the interaction between an impacting fine particle and a single static bed particle. In this setup, a fine particle is released from rest and allowed to fall under gravity from an initial height of $2 d_f \approx 0.29 d_l$ above the contact point on
 the large particle, reaching an impact velocity of $v_0 = 0.15\,\mathrm{m/s}$. The initial position of the fine particle is offset by $0.5d_l$ from a vertical line through the large particle so that the fine particle does not directly impact the top of the large particle. We vary the Bond number $Bo$, sliding friction coefficient $\mu$, and restitution coefficient $e_n$. Four distinct post-impact behaviors are observed as illustrated in Fig.~\ref{fig:two_particle_interactions}, which shows time-series of the fine-particle position (colormap).

At a relatively large restitution of $e_n=0.8$, the fine particle rebounds after colliding with the large particle, as illustrated in Fig.~\ref{fig:two_particle_interactions}(a). When the restitution coefficient $e_n$ is reduced to $0.2$, the particle no longer rebounds but adheres to the surface and slides a small distance before stopping, as shown in Fig.~\ref{fig:two_particle_interactions}(b). When the sliding friction is reduced from 0.3 to 0.02 while maintaining the adhesion condition associated with small $e_n$, the particle slides along the surface of the large particle past its lowest point and then back before eventually stopping just shy of the lowest point on the large particle, as shown in Fig.~\ref{fig:two_particle_interactions}(c). However, when the cohesion is reduced from $Bo=8$ to $Bo=3$, the fine particle slides down the side of the large particle after the initial impact but then detaches, as shown in Fig.~\ref{fig:two_particle_interactions}(d). In the bed, case (a) represents rebounding behavior, while cases (b - d) represent adhesion. In cases (b) and (c), cohesion prevents further percolation of a fine particle, while cases (a) and (d) allow continued percolation through the bed.

These observations suggest that trapping in a cohesive system involves two sequential processes. First, the fine particle collides with the large particle and either rebounds or adheres. If the particle adheres after a collision, it moves along the surface and is either trapped or detaches. Therefore, the adhesion condition is a necessary but not sufficient requirement for trapping. Trapping is strongly influenced by both friction and cohesion. In the following sections, we explore the impact of these two mechanisms on particle interactions. Note that other means of trapping fine particles can occur in addition to cohesive collisional trapping including geometric trapping where the fine particle is too large to pass through a pore throat~\cite{Gao2023} and frictional trapping where the fine particle is trapped due to friction between two large particles~\cite{Vyas2025a}.

\subsection{Adhesion Condition}

\subsubsection{Effective restitution coefficient and critical velocity}
When a particle contact is cohesive, the outcome of a collision depends on the relative impact velocity. For a normal collision between two particles with the particle contact colinear with the centers of the particles (no tangential velocity), the equation of motion in terms of the virtual overlap, $\delta$, is given by Newton's second law:

\begin{equation}\label{eq:NewtonODE}
\begin{aligned}
m_{\text{eff}} \ddot{\delta} =\;&
- \frac{4}{3} E_{\text{eff}} r_{\text{eff}}^{1/2} \delta^{3/2}
+ 4 \pi \sigma r_{\text{eff}}
+ C \delta^{1/4}\dot{\delta}, \\
\text{where}\quad
C =\;&
-2 \sqrt{\frac{5}{6}}
\frac{\log(e_n)}{\sqrt{\pi^2 + \left(\log(e_n)\right)^2}}
\sqrt{2 E_{\text{eff}} r_{\text{eff}}^{1/2} m_{\text{eff}}}.
\end{aligned}
\end{equation}
The first term represents the Hertzian elastic force~\cite{Thornton2011}, the second term corresponds to the constant cohesive force from the DMT model~\cite{Derjaguin1975}, and the third term accounts for viscous damping for a constant normal restitution coefficient, $e_n$, for non-cohesive particles~\cite{Brilliantov1996,Vyas2025a}. Here, $m_{\text{eff}}$ is the effective mass of the two colliding particles, $m_{\text{eff}} = m_l m_f /(m_l + m_f)$. The presence of the cohesive force renders the equation inhomogeneous. Its positive contribution allows for the possibility of a positive stable focus in the velocity-displacement phase space, which can result in adhesion for low enough collision velocity.

To solve Eq.~\eqref{eq:NewtonODE}, we impose the initial conditions $\delta(0) = 0$ and $\dot{\delta}(0) = v_0$ at the moment of initial impact. Eq.~\eqref{eq:NewtonODE} describes the increasing deformation at the point of contact and then the subsequent decreasing deformation as the particles move away from each other. The time at separation, $t_0$, is defined as the first instance when $\delta(t_0) = 0$ at the end of the particle contact, noting that $\delta(t) > 0$ for all $t < t_0$, consistent with the DMT separation criterion. We introduce the effective restitution coefficient as~\cite{Abbasfard2016,Murphy2017}
\begin{equation} \label{eq:effective_restitution}
    e_{\text{eff}} = -\frac{\dot{\delta}(t_0)}{\dot{\delta}(0)}.
\end{equation}
If $e_{\text{eff}} > 0$, the particles rebound with a finite velocity, $\dot{\delta}(t_0)$. If the cohesive force and damping are large enough, the particles do not separate, indicating adhesion for which $e_{\text{eff}} = 0$.

\begin{figure}[htbp]
  \centering
  \includegraphics[width=1\linewidth]{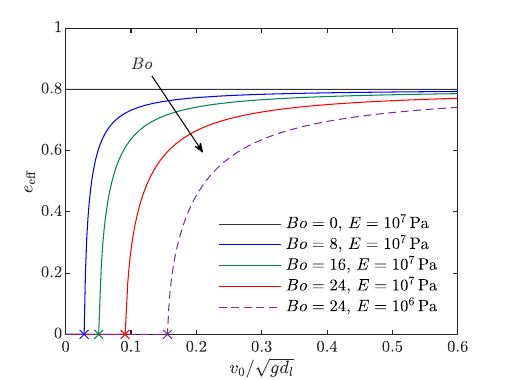}
  \caption{Effective restitution coefficient, $e_{\text{eff}}$, vs scaled normal impact velocity, $v_0/\sqrt{gd_l}$, for different $Bo$ and $e_n=0.8$. Critical velocities $v_{\text{cr}}$ are denoted by $\times$ markers.}
  \label{fig:effective_restitution_coefficient}
\end{figure}

We vary the impact velocity $v_0$, scaled by the characteristic gravity-driven percolation velocity, $\sqrt{gd_l}$, and compute $e_{\text{eff}}$ under different Bond numbers ($Bo$), as shown in Fig.~\ref{fig:effective_restitution_coefficient} for an example of a relatively soft material $E=10^7\,\mathrm{Pa}$ with $R=7$ and $e_n=0.8$. For $Bo = 0$, $e_{\text{eff}}$ remains constant and equal to $e_n$, independent of impact velocity. For $Bo > 0$, $e_{\text{eff}}$ is zero at low impact velocities, indicating adhesion (no rebound). As $v_0$ increases beyond a critical velocity, $v_{\text{cr}} = \min \left\{ v_0 \, \big| \, e_{\text{eff}}(v_0) > 0 \right\}$, which depends on $Bo$ and $E$ for a given $e_n$, $e_{\text{eff}}$ becomes nonzero, signifying rebound and separation. Clearly, $e_{\text{eff}}$ strongly depends on cohesion at low $v_0$, but at sufficiently high $v_0$, $e_{\text{eff}}$ asymptotically approaches $e_n$, regardless of $Bo$. Reducing Young’s modulus under identical conditions shifts the transition to higher velocities, reflecting an increased critical impact velocity required for rebound due to the increase in viscous dissipation associated with greater deformation. 

Physically, $v_{\text{cr}}$ marks the threshold between adhesion and rebound. For $v_0 < v_{\text{cr}}$, cohesive and dissipation forces dominate and the particles remain adhered after collision. For $v_0 > v_{\text{cr}}$, the kinetic energy is sufficient to overcome cohesion, resulting in rebound. $v_{\text{cr}}$ increases with $Bo$, reflecting stronger cohesive interactions. This behavior is consistent with similar findings for cohesive van der Waals forces~\cite{Kellogg2017,Wang2025}.

\begin{figure}[htbp]
  \centering
  \includegraphics[width=1\linewidth]{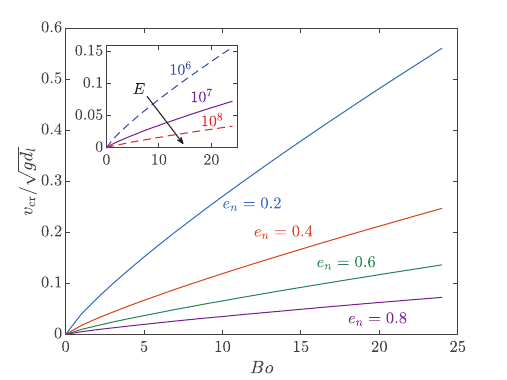}
  \caption{Critical velocity of normal collision vs Bond number for different $e_n$, with inset showing effect of different $E$ for $e_n=0.8$.}
  \label{fig:critical_velocity}
\end{figure}

The critical velocity $v_{\text{cr}}$ based on solving Eq.~\eqref{eq:NewtonODE} is plotted as a function of $Bo$ for various $e_n$ in Fig.~\ref{fig:critical_velocity}. The critical velocity increases monotonically with increasing $Bo$, decreasing $e_n$, and decreasing $E$ (see inset), reflecting how both cohesive forces and particle properties contribute to $v_{\text{cr}}$. Particles are less likely to rebound upon contact when cohesion is stronger or when $e_n$ or $E$ are smaller, requiring greater impact velocity to initiate separation. Specifically, both $e_n$ and $E$ directly influence the dissipative forces reflected in the last term in Eq.~\eqref{eq:NewtonODE}. Returning to Fig.~\ref{fig:two_particle_interactions}, the dependence of $v_{\mathrm{cr}}$ on $e_n$ and $Bo$ defines whether the impacting fine particle rebounds, as in Fig.~\ref{fig:two_particle_interactions}(a), or remains attached to the large particle, as in the other cases.

\subsubsection{Adhesion probability and collision frequency}

To isolate the role of collisions in the trapping process, we first consider an idealized case of purely adhesive trapping (Fig.~\ref{fig:two_particle_interactions}(b, c)) where the fine particle does not rebound (Fig.~\ref{fig:two_particle_interactions}(a)) or slide along the surface and eventually detach (Fig.~\ref{fig:two_particle_interactions}(d)). This serves as an asymptotic limit for systems characterized by strong cohesive and frictional interactions. In this simplified picture, fine particles interact with large particles through collisions, and either stick to its surface without sliding or rebound. Once a fine particle sticks, it stops percolating.

Under these assumptions, fine particle trapping depends on two factors: the probability of trapping upon collision and the frequency of collisions. Each percolating fine particle possesses an initial velocity prior to collision. Since gravity acts exclusively in the vertical direction, free-falling particles experience no horizontal forces prior to collision and therefore exhibit no horizontal acceleration. Hence, the horizontal velocity can be used as a proxy for the characteristic normal impact velocity. While this approach slightly overestimates the true impact velocity due to non-normal collisions, it preserves the correct scaling behavior and trend.

\begin{figure}[htbp]
  \centering
  \includegraphics[width=1\linewidth]{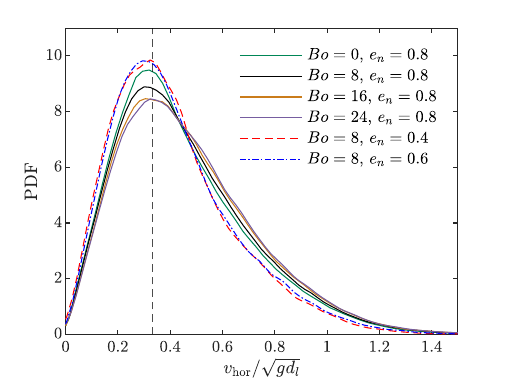}
  \caption{Probability density function of percolating free fine particle horizontal velocity with peak at $v_{\text{hor}}/\sqrt{gd_l}\approx 1/3$ (vertical dashed line) for different $Bo$ and $e_n$ with $\mu=0.05$.}
  \label{fig:horizontal_distribution}
\end{figure}

Figure~\ref{fig:horizontal_distribution} presents the measured distribution of horizontal velocities for fine particles percolating through the bed of large particles under various conditions. At a fixed restitution level (solid curves), increasing $Bo$ slightly reduces the peak and slightly broadens the distribution; for sufficiently large $Bo$, the PDFs collapse onto a common curve. For a given cohesion level, decreasing $e_n$ leads to a slightly narrower and taller peak. Despite these minor differences, all curves exhibit a similar shape with a peak located at $v_{\mathrm{hor}}/\sqrt{g d_l} \approx 1/3$, a comparatively small value relative to the characteristic vertical velocity. This indicates that fine particles undergo frequent collisions that rapidly dissipate horizontal momentum as they migrate downward through the bed.

\begin{figure}[htbp]
  \centering
  \includegraphics[width=1\linewidth]{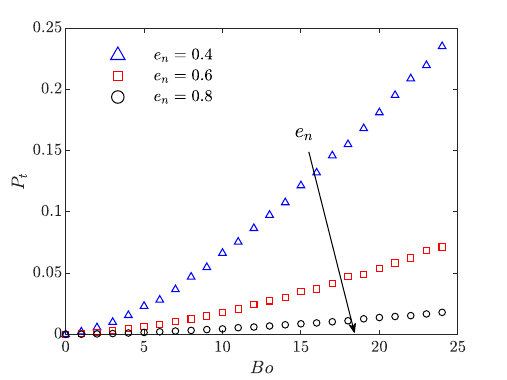}
  \caption{Trapping probability per collision based on $v_{\text{hor}}$ vs $Bo$ for various $e_n$ with $\mu=0.05$.}
  \label{fig:trapping_probability}
\end{figure}

The trapping probability per collision is the fraction of collisions where $v_{\mathrm{hor}} < v_{\mathrm{cr}}$, which can be expressed as
\begin{equation} \label{eq:trapping_probablity}
    P_t(Bo,\,e_n) = \int_0^{v_{\text{cr}}(Bo,\,e_n)} \text{PDF}(v_{\text{hor}}) \, dv_{\text{hor}}.
\end{equation}
With respect to Fig.~\ref{fig:horizontal_distribution}, this corresponds to the area under the curve to the left of the critical velocity, $v_{\mathrm{cr}}$, which depends on $Bo$ and $e_n$ (see Fig.~\ref{fig:critical_velocity}). As shown in Fig.~\ref{fig:trapping_probability} for the representative conditions considered here, the trapping probability $P_t$ is strongly dependent on $e_n$, reflecting the corresponding increase in $v_{\mathrm{cr}}$ for decreasing $e_n$.  The trapping probability also increases monotonically with $Bo$ for fixed $e_n$. This trend follows directly from the monotonic increase of the critical velocity $v_{\mathrm{cr}}$ with $Bo$, evident in Fig.~\ref{fig:critical_velocity}.

\begin{figure}[htbp]
  \centering
  \includegraphics[width=1\linewidth]{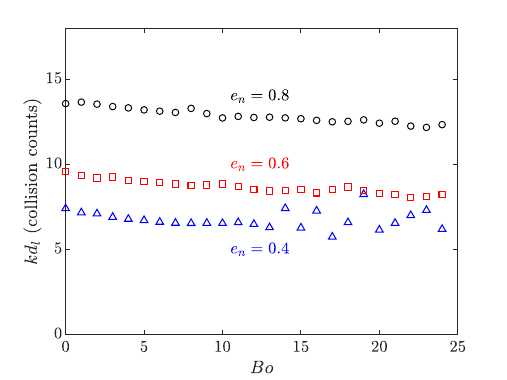}
  \caption{Number of collisions, $kd_l$, for a fine particle percolating a distance $d_l$ vs $Bo$ for $e_n = 0.4$, $0.6$, and $0.8$, and $\mu=0.05$.}
  \label{fig:collision_counts}
\end{figure}

To estimate the trapping distance, we also quantify the spatial collision frequency of fine particles with bed particles, $k$, by recording a collision event whenever a fine particle transitions from a free state to being in contact with a bed particle. This requires storing data at a higher time resolution than the usual simulations to ensure that collisions are adequately resolved. Figure~\ref{fig:collision_counts} shows the number of collisions $kd_l$ occurring over distance $d_l$ in the static bed. The collision frequency depends strongly on the restitution coefficient because higher values of $e_n$ lead to greater velocity fluctuations for the fine particles~$e_n$~\cite{Vyas2025b} and more frequent collisions. The dependence of $k$ on $Bo$ is minimal.

\subsubsection{Connecting collisions with the trapping distance}
Consider now the collision of a cohesive fine particle with a bed particle. The fine particle sticks to the bed particle with trapping probability $P_t$ and rebounds from the bed particle with probability $1 - P_t$. After $n$ collisions, the probability that a fine particle is still free (not yet trapped) is $P_f = (1 - P_t)^n$ assuming independent collisions. Before any collisions $(n=0)$, $P_f(n = 0) = 1$, and, since $0\leq P_t \leq 1$, $P_f(n \rightarrow \infty) = 0$, indicating that the particle will eventually be trapped if $P_t>0$. The number of collisions, $n$, can be expressed as $n = -k\Delta z$, noting that $\Delta z = z - z_0$ is negative. The probability of a fine particle remain untrapped after descending a distance $\Delta z$ is
\begin{equation}\label{eq:total_collision}
P_f\left(\Delta z\right) = \left(1-P_t\right)^{-k\Delta z},
\end{equation}
From the perspective of the characteristic trapping depth, this same probability can be expressed using Eq.~\eqref{eq:exponential} as
\begin{equation}\label{eq:total_statistics}
    P_f^*\left(\Delta z\right)=\int_{-\infty}^{\Delta z} \frac{1}{\lambda} e^{\Delta z^*/\lambda} d\Delta z^* = e^{\Delta z/\lambda},
\end{equation}
where $\lambda$ is the characteristic trapping distance.

Equating Eqs.~\eqref{eq:total_collision} and \eqref{eq:total_statistics} gives
\begin{equation}\label{eq:collision_model}
    \frac{1}{\lambda} 
    = -k\ln(1 - P_t).
\end{equation}
This expression directly links the macroscopic trapping distance description and the particle-level collision dynamics. The left-hand side represents the macroscale behavior obtained by fitting the trapped particle distribution to the exponential form in Eq.~\eqref{eq:exponential}, and therefore incorporates the combined influence of all parameters affecting the trapping distance, including $Bo$, $e_n$, and $E$. The right-hand side, in contrast, follows from particle-resolved collision behavior. The parameter $k$ characterizes the collision frequency, which increases with $e_n$ and remains nearly independent of $Bo$, while $P_t$ represents the trapping probability per collision governed primarily by the critical velocity via Eq.~\eqref{eq:trapping_probablity}. Since each side of Eq.~\eqref{eq:collision_model} arises from fundamentally different scales but yields a consistent mapping, this result demonstrates that the framework is not limited to a particular set of parameters but remains valid across a broad range of conditions.

\begin{figure}[htbp]
  \centering
  \includegraphics[width=1\linewidth]{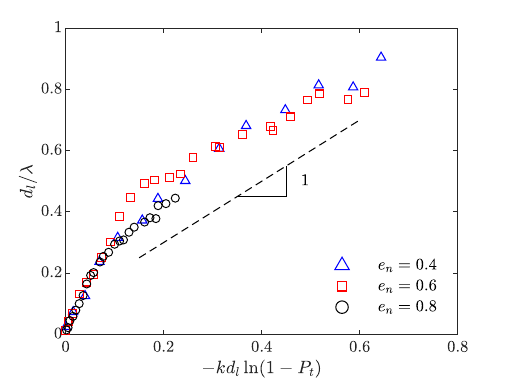}
  \caption{Correlation between the inverse trapping distance $d_l/\lambda$ and the collisional prediction for trapping distance $-kd_l\ln(1-P_t)$ for $e_n = 0.4$, $0.6$, and $0.8$, $\mu=0.05$, and Bond numbers $Bo$ ranging from $0$ to $24$.}
  \label{fig:collision_statistics_compare}
\end{figure}

We verify the proposed relationship using simulation data for $\mu=0.05$ and examine how the particle-scale parameters map onto the macroscopic trapping distance. Figure~\ref{fig:collision_statistics_compare} shows the correlation between the inverse trapping distance $d_l/\lambda$ and the collisional prediction $-kd_l\ln(1-P_t)$ for restitution coefficients $e_n = 0.4,\,0.6,\,0.8$. For each restitution coefficient, the Bond number increases from left to right in the figure, spanning $Bo$ from 0 to 24 for $e_n = 0.6$ and $0.8$, and from 0 to 12 for $e_n = 0.4$. For sufficiently large $Bo$, the system is in a collision-dominated regime, and $d_l/\lambda$ depends linearly on $-kd_l\ln(1-P_t)$ with a slope of approximately 1 across all restitution levels. In this regime, the macroscopic trapping distance is therefore captured directly by the particle-scale collision statistics. 
At small $Bo$, the slope is steeper because sliding friction and cohesion act simultaneously such that the trapping behavior is no longer governed purely by the collisional processes. Despite these deviations when sticking probability is low (i.e. $P_t\rightarrow 0$ as $\lambda\rightarrow\infty$), the overall trend demonstrates that the inverse trapping distance can be approximated from particle-level interactions. The agreement across a wide range of cohesion strengths and restitution coefficients confirms that the collisional framework is broadly applicable. 

\subsection{Frictional Effects}\label{sec:friction}

In the previous section, we used constant friction to focus on the effects of cohesion. However, when a fine particle does not rebound, friction is enhanced by the increased normal force due to cohesive attraction. The effective normal force in such cases is given by~\cite{Marshall2009}
\begin{equation} \label{eq:enhanced_normal}
    F_{n0} = \left\lVert \mathbf{F}_{\mathrm{ne}} \cdot \hat{\mathbf{n}} + 2F_c \right\rVert,
\end{equation}
where $\mathbf{F}_{\mathrm{ne}}$ denotes the net external force (e.g., from the fine particle weight) at the contact point, which is equal and opposite to the repulsive force ($F_{\text{DMT}}$ for the simulations here), $\hat{\mathbf{n}}$ is the unit normal vector at the contact point, and $F_c$ is the pull-off force arising from cohesion. The coefficient 2 ensures that the tangential force is $\mu F_c$ when the particles are about to separate.

For a fine particle sliding along the surface of a large spherical particle with negligible centripetal acceleration, such as Fig.~\ref{fig:two_particle_interactions}(c, d), the normal contact force is the sum of the normal component of the fine particle's weight and the cohesive contact force. Hence, the frictional force resisting motion in the tangential direction is expressed as
\begin{equation} \label{eq:friction_on_sphere}
    f = \mu m_f g \left( \cos\theta + 2Bo \right), \quad \theta \in [0, 180^\circ],
\end{equation}
where $\theta$ is the angular position measured from the top of the large particle. Note that gravity increases the friction force when the fine particle is contacting the upper hemisphere of the large particle and reduces the friction force on the lower hemisphere.

For the simplest case, to determine the condition under which a fine particle remains stationary on the surface of a large particle, the frictional force balances the component of the fine particle weight in the tangential direction. The critical friction coefficient required to prevent sliding depends on $\theta$:
\begin{equation} \label{eq:critical_friction}
    \mu_c = \max_{0 \le \theta \le 180^\circ} \left( \frac{\sin\theta}{\cos\theta + 2Bo} \right),
\end{equation}
which attains its maximum at $\theta_{max} = 2\arctan\left( \sqrt{\frac{2Bo + 1}{2Bo - 1}} \right)$, corresponding to a location slightly below the equator. This point represents the angular position where the tangential component of gravity requires the largest coefficient of friction to prevent sliding.

Substituting $\theta_{max}$ into Eq.~\eqref{eq:critical_friction} defines the criterion for preventing sliding of a stationary fine at any location:
\begin{equation} \label{eq:static_equilibrium_condition}
    \mu\sqrt{4Bo^2 - 1} \geq 1.
\end{equation}
Combinations of $\mu$ and $Bo$ for which the LHS of Eq.~\eqref{eq:static_equilibrium_condition} exceeds one ensure that the fine particle remains stationary at all positions on the large particle under the combined influence of gravity, frictional forces, and cohesive forces.

This condition has significant physical implications. If the combined effect of sliding friction and cohesion is below the threshold defined by Eq.~\eqref{eq:static_equilibrium_condition}, a fine particle that sticks to the surface without rebounding can slide down the surface of the large particle (Fig.~\ref{fig:two_particle_interactions}(c, d)) and potentially detach (Fig.~\ref{fig:two_particle_interactions}(d)). Conversely, if sliding friction exceeds this threshold, the particle is more likely to remain fixed upon contact or slide only a short distance (Fig.~\ref{fig:two_particle_interactions}(b)), depending on the tangential velocity at impact. Thus, the static equilibrium condition serves as a predictive criterion for non-rebounding post-collision particle behavior. A similar criterion can be derived to account for the effects of rolling resistance. 

\begin{figure}[htbp]
  \centering
  \includegraphics[width=1\linewidth]{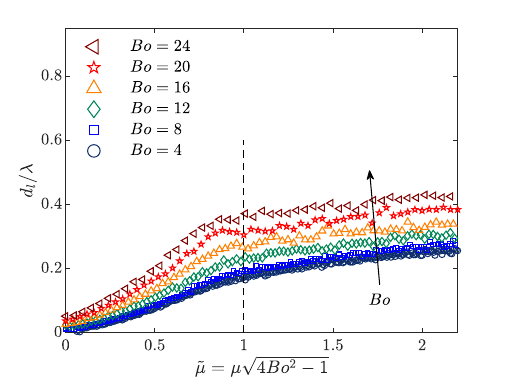}
  \caption{Inverse trapping distance, $d_l/\lambda$, vs rescaled
sliding friction coefficient, $\tilde{\mu}=\mu \sqrt {4 Bo^2 -1}$.}
  \label{fig:critical_friction}
\end{figure}

Returning to Fig.~\ref{fig:trapping_length_friction}, the inverse trapping distance exhibits a kink in its dependence on the Bond number. The static equilibrium condition of Eq.~\eqref{eq:static_equilibrium_condition} and its physical interpretation allow us to predict the location of this kink. Figure~\ref{fig:critical_friction} presents the same data  as in Fig.~\ref{fig:trapping_length_friction} for the inverse trapping distance but as a function of the rescaled friction coefficient including the effect of cohesion: $\tilde{\mu}=\mu \sqrt {4 Bo^2 -1}$. Notably, the kink consistently appears near $\tilde{\mu}=1$. When $\tilde{\mu} > 1$,  $d_l/\lambda$ increases slowly with increasing $\tilde{\mu}$ and with a slope that is independent of $Bo$.  This implies that the $Bo$-dependent offset in the curves is entirely due to the increase in trapping probability associated with the increase in $v_\mathrm{cr}$ with increasing $Bo$ (see Fig.~\ref{fig:critical_velocity}); once adhered to the surface, the resulting average sliding behavior is identical (i.e., the probability that a fine is trapped or detaches depends only on $\tilde{\mu}$ and not $Bo$). The slow increase in $d_l/\lambda$ with increasing $\tilde{\mu}$ represents the decreasing fraction of post-adhesion contact locations and tangential velocities that subsequently detach.  

In contrast, when $\tilde{\mu} < 1$, particles are more likely to slide along the surface and detach as cohesive forces are insufficient to counteract gravitational and centripetal accelerations. This leads to reduced trapping and a stronger dependence of the trapping distance on $\mu$ and $Bo$ across all cohesion levels. Unlike the case for $\tilde{\mu}>1,$ the slope of $d_l/\lambda$ vs $\tilde{\mu}$ increases with increasing $Bo$ for any fixed value of $\tilde{\mu}<1$.

\subsection{Double contact}

\begin{figure}[htbp]
  \centering
  \includegraphics[width=1\linewidth]{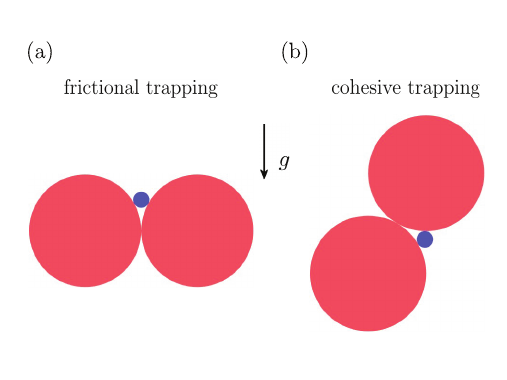}
  \caption{Examples of double contact trapping under gravity for $R=7$: (a) frictional trapping; (b) cohesive trapping. }
  \label{fig:double_contact}
\end{figure}

An intriguing observation is that the curves in Fig.~\ref{fig:critical_friction} start at relatively low values for $d_l/\lambda$ (deep penetration) as $\tilde{\mu} \rightarrow 0$, even at high Bond numbers, suggesting that particles may still detach despite strong cohesion, which seems counterintuitive. One possible mechanism involves double contact cases, illustrated in Fig.~\ref{fig:double_contact}. 

In Fig.~\ref{fig:double_contact}(a), the fine particle becomes trapped through a frictional double contact configuration in which it touches two large particles simultaneously and forms a mechanically stable geometry that resists sliding down the valley between the large particles (normal to the page)~\cite{Vyas2025a}. Figure~\ref{fig:double_contact}(b) illustrates a cohesive double contact configuration. In this case, cohesive forces at both contact points act together to hold the fine particle in place. As the load is shared between two contacts, certain geometric arrangements permit a fine particle to be trapped with even lower cohesion than required for single contact trapping (Fig.~\ref{fig:two_particle_interactions}(b, c)). 

\begin{figure}[htbp]
  \centering
  \includegraphics[width=1\linewidth]{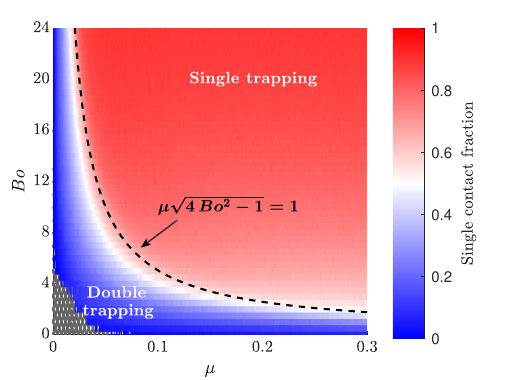}
  \caption{Phase diagram of single contact fraction showing fraction of all trapped particles with a single contact. Hashed pattern indicates conditions under which an insufficient number of fines are trapped to obtain reliable statistics.}
  \label{fig:double_contact_phase_chart}
\end{figure}

As fine particles slide along the surfaces of larger particles, double contacts may either promote trapping or lead to detachment from the original particle, depending on the local geometry and loading conditions. Figure~\ref{fig:double_contact_phase_chart} presents the phase diagram for the fraction of trapped fine particles that have a single contact with any large bed particle, while the remainder have at least two contacts. In the upper-right region, where both friction and cohesion are strong, most trapped particles are immobilized through a single contact configuration. As friction or cohesion is decreased, the single contact fraction declines, indicating that double contact trapping becomes increasingly dominant.

Plotting the static equilibrium condition given by Eq.~\eqref{eq:static_equilibrium_condition} on the phase diagram effectively separates the regions dominated by single contact and double contact trapping. When the rescaled friction satisfies $\mu\sqrt{4Bo^{2}-1} > 1$ (above the curve in Fig.~\ref{fig:double_contact_phase_chart}), particles tend to stick after adhering upon collision with little subsequent sliding. As a result, they are unlikely to encounter a second large particle and are therefore more likely to be trapped via a single contact. In contrast, when the rescaled friction falls below 1, particles are more likely to slide along the surface after sticking. In this regime, apart from the possibility of detachment, particles frequently make contact with a neighboring large particle, leading to double contact trapping.

When cohesion is strong, detachment from a single large particle becomes highly unlikely because cohesive forces dominate. However, in certain cases detachment can occur through a double contact mechanism, where a fine particle slides down the surface of a large particle until it collides with a second large particle while still in contact with the first. The collision with the second large particle may result in a rebound adequate to overcome the cohesive force binding the fine particle to the first surface. As a result, the fine particle may detach from the original particle even with strong cohesion.

\section{Conclusions} \label{sec:Conclusions}

Percolation of fine particles in a static bed of larger particles is central to many industrial and natural processes. Fines either pass freely through the bed or become trapped depending on multiple factors including the particle sizes, friction and restitution coefficients, polydispersity, and cohesion. Unlike previous studies that consider non-cohesive systems, we examine how cohesion, restitution, and friction combine to influence percolation even when the particle size ratio geometrically permits free passage through pore throats. Cohesion is characterized using the Bond number $Bo$, and simulations are performed with the DMT cohesion model, although other cohesive contact models produce similar results.

Strong cohesion, particularly when combined with friction, induces non-geometric trapping when the system would otherwise be free sifting (Fig.~\ref{fig:passing_and_trapping}). As cohesion increases, fines are trapped at shallower depths (Fig.~\ref{fig:percolattion}). The spatial distribution of trapped particles is well described by an exponential form (Eq.~\ref{eq:exponential}), and the characteristic trapping distance provides a metric for percolation behavior. The inverse trapping distance increases monotonically with $Bo$, indicating that the trapping distance decreases with $Bo$ (Fig.~\ref{fig:trapping_length_Bo}). For all cohesion levels, the inverse trapping distance also grows with friction coefficient $\mu$, with a kink separating two distinct trapping regimes below which friction dominates and above which cohesion dominates (Fig.~\ref{fig:trapping_length_friction}). A phase diagram in the $(Bo,\mu)$ plane (Fig.~\ref{fig:phase_chart}) shows that continuous families of parameters lead to identical trapping distance. Stiffness further modifies percolation dynamics as decreasing $E$ enhances trapping (Fig.~\ref{fig:trapping_length_YoungsM}).

To interpret these trends, we analyze interactions between a fine particle and a single bed particle (Fig.~\ref{fig:two_particle_interactions}) to decompose the trapping process into two steps: (i) collision (rebound or adhere) and (ii) subsequent motion of the fine particle sliding along the surface of the large particle if it does not rebound. In the first step, the coupled effects of cohesion, restitution coefficient, and impact velocity determine whether rebounding occurs. Fines may fail to rebound after a collision due to large cohesion or low restitution or Young's modulus. We propose a collisional model that incorporates a trapping probability per collision (Fig.~\ref{fig:trapping_probability}) and a collision frequency (Fig.~\ref{fig:collision_counts}) to predict the trapping distance in the collision-dominated regime (Fig.~\ref{fig:collision_statistics_compare}). For non-rebounding collisions, frictional effects are enhanced by cohesion and, when large enough, prevent the fine particle from subsequently detaching. A static equilibrium condition (Eq.~\eqref{eq:static_equilibrium_condition}) based on force balance predicts whether a fine particle remains stationary after contact. Thus, post-collision motion -- sliding, detachment, or arrest -- is controlled by the interplay of friction and cohesion. Rescaling friction by this condition aligns the kinks in the trapping distance data (Fig.~\ref{fig:critical_friction}), marking the transition from both friction and cohesion playing roles to cohesion dominating. This framework also distinguishes single contact from double contact trapping (Fig.~\ref{fig:double_contact_phase_chart}), indicating when a fine particle is likely to form a second contact after a non-rebounding collision.

In addition to clarifying the mechanisms in percolation of non-interacting cohesive fines in static beds, these results also provide a framework that can be extended to sheared granular flows. More broadly, they show that percolation is not governed by geometric accessibility alone, but by particle-scale interaction dynamics that can override geometric expectations. In such flows, the dynamics of how fine particles encounter and interact with large particles and the conditions under which they become temporarily trapped or continue percolating remain central questions. While cohesion plays a major role in promoting trapping in static beds, shear introduces continuous particle rearrangements that can dislodge fine particles, enabling further percolation rather than permanent immobilization. This suggests that in flowing systems, trapping and release may form a dynamic balance. These considerations highlight that cohesive fine particle percolation in shear flows is considerably more complex than the static case examined here and deserves further investigation.

\begin{acknowledgments}
We thank Dhairya R. Vyas for insightful discussions. This material is based upon work supported by the National Science Foundation under Grant No.CBET-2429545, and by the International Fine Particle Research Institute.
\end{acknowledgments}


\bibliography{Reference}

\end{document}